\documentstyle[11pt,newpasp,epsf,epsfig]{article}

\def\solm{M$_{\odot}\,$}
\def\kms{km s$^{-1}\,$}
\def\etal{{\it et~al.\ }}

\begin{document}

\title{Simulations of Gas Flow from a Galactic Disk \\ to the BH-dominated
Central Disk}

\author{Witold Maciejewski}
\affil{Theoretical Physics, University of Oxford,\\ 
1 Keble Road, Oxford OX1 3NP, United Kingdom,\\ 
and Obserwatorium Astronomiczne Uniwersytetu Jagiello\'nskiego}

\begin{abstract}
With a help of hydrodynamical models, we explore fundamental modes of
gas flow in central parts of a galaxy potential with non-axisymmetric
perturbations of various strengths.  Our grid-based algorithm allows
us to get a detailed picture of gas dynamics down to about 10 pc from
the galaxy center, where the central black hole becomes dominant.  Gas
inflow in a bar with an Inner Lindblad Resonance often stops at the
nuclear ring.  We find that embedded bars in resonant coupling
are unlikely to increase the inflow, but for certain gas parameters in
a single bar, inflow to the center develops in a spiral shock,
similar to the patterns observed in nuclear discs of real galaxies.  
The spiral pattern persists for very weak non-axisymmetries of the 
potential.
\end{abstract}

\section{Introduction}
Interstellar gas falls towards the center of a galaxy if torques act
on it and remove its angular momentum.  On scales larger than several
parsecs, gravitational torques from non-axisymmetric mass distribution
in stars dominate.  They can be caused by external factors
(interactions and mergers) or internal ones (bars and other results of
disk instabilities).  Here, we focus on gas flow in barred galactic
potentials: if the bar is strong enough, the $x_1$ orbits aligned with
it develop cusps at their apocenters, and gas clouds there
run onto each other, causing shocks.  The shocked gas loses angular
momentum to the bar, and a considerable inflow occurs at radii where
shocks are present.  In a bar with an Inner Lindblad Resonance (ILR),
orbiting gas shifts gradually from $x_1$ to $x_2$ orbits, and
eventually settles in a nuclear ring, located approximately at the
bar's ILR, at a radius of order 1 kpc.

Accumulation of gas in the nuclear ring may trigger a nuclear
starburst, but Active Galactic Nuclei (AGN) require mass accretion to
within a few parsecs of the galaxy center.  A mechanism of bars within
bars has been proposed to drive further inflow, and thus feed the AGN
in a manner similar to the inflow on large scales (Shlosman, Frank \&
Begelman 1989).  In its original form, it involved a cascade of
nested, independently rotating gaseous bars, with the inner bar
forming as a result of dynamical instability in the gaseous nuclear
disk. Such instability develops rapidly, and may cause a significant
inflow, but one should not expect well-ordered flow morphology there.
On the other hand, bars within bars observed in the stellar component
of galaxies are likely to be of different nature: even if they form as
an effect of gravitational instability, their motion is well-ordered,
and their lifetime longer.

Here we present some preliminary results on fundamental modes of gas flow
in strong single and double bars, and in weak oval distortions, and we
examine how efficient they can be in feeding the nucleus. We give a
comprehensive account of this work elsewhere (Maciejewski \etal 2001).

\section{Numerical code, potential and initial conditions}
We used the grid-based Eulerian hydrodynamical code CMHOG, written by
James M. Stone at the University of Maryland and adopted to the polar
grid in two dimensions, with isothermal equation of state, by Piner
\etal (1995).  The code implements the piecewise parabolic method (PPM), 
and the use of
square cells gives high resolution at the center: up to 0.4 pc at the
radius of 20 pc.  Self-gravity of gas is neglected, and the numerical
viscosity is very low --- the infall to the center is negligible in
an axisymmetric potential for timescales considered here.

The grid extends from 20 or 100 pc to 16 kpc in radius, and the gas
flow is followed in a potential of Model 2 of a doubly barred galaxy 
constructed by Maciejewski \& Sparke (2000). Azimuthal averaging of 
the secondary bar gives our potential of a singly-barred galaxy.
The semi-major axis of the main bar is 6
kpc with a rotation period of 0.171 Gyr; the corresponding values for
the secondary bar are 1.2 kpc and 0.056 Gyr.  Gas is initially
distributed uniformly over the grid, and forced to rotate around the
center with the circular velocity in the azimuthally averaged
potential.  Then the outer bar of Model 2 is introduced by gradual
transfer of mass from the bulge.  In runs with double bars, the
secondary bar is introduced in the same way, after the flow in the
primary bar has stabilized.

\section{Gas inflow in single and double bar compared}
General features of gas flow in a single strong, fast-rotating bar
with two ILRs can be seen in Fig.1, with shocks showing up best in the
square of velocity divergence in the gas (div$^2{\bf v}$, div {\bf v}
$<0$).  Gas loses angular momentum in the principal shocks (PS) along
the leading sides of the bar and falls towards the center. Inside the
ILR (i.e. between the inner and outer ILR)
a gaseous nuclear ring (NR) forms, and the inflow stagnates there,
with star formation likely.  A weak spiral density wave propagates
inwards from the nuclear ring, but the rate of gas inflow to the
center is very small: less than $10^{-4}$ \solm/yr.  

\begin{figure}[t]
\vspace{-4cm}
\plotfiddle{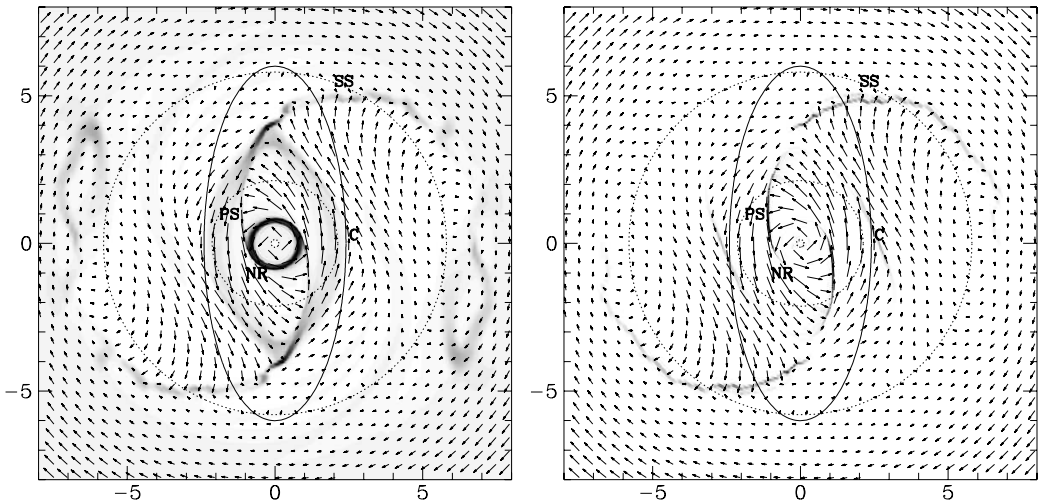}{6cm}{0}{66}{66}{-200}{-470}
\vspace{3.6cm}
\caption{Density ({\it left}), and $div^2\bf v$ ({\it right}) snapshots of
gas flow in a single bar at time 1 Gyr. The bar is outlined in solid,
and the dotted curves mark the inner ILR at 0.13 kpc, the outer
ILR at 2.14 kpc, and corotation at 7.78 kpc. Gas velocity in
the frame rotating with the bar is marked by arrows. Positions of the
principal shock (PS), nuclear ring (NR), spiral shock (SS), and the
convergence region (C) are marked. Units on axes are in kpc.}
\end{figure}

By extending the
concept of multiple gaseous bars proposed by Shlosman et al.\ (1989)
to the stellar bars, it is thought that the secondary bar may force
gas deeper into the potential well, if gas flow in the secondary bar
is analogous to the one above.
To verify this view, Maciejewski \& Sparke (2000) constructed a model
of a dynamically possible doubly barred galaxy.  They started with the
potentials of two rigid, independently rotating bars embedded in a
disk and halo, and searched for particle distributions that could
support the assumed potential.  Despite the absence of a conserved
Jacobi integral in this system, the orbits are not mostly chaotic.
Particle motions appear well-ordered, if one looks not at orbits, but
at sets of particles forming closed curves which return to their 
original position after two bars realign. These sets (called here loops)
are equivalent to orbits in a time-independent potential.

\begin{figure}[t]
\vspace{-4cm}
\plotfiddle{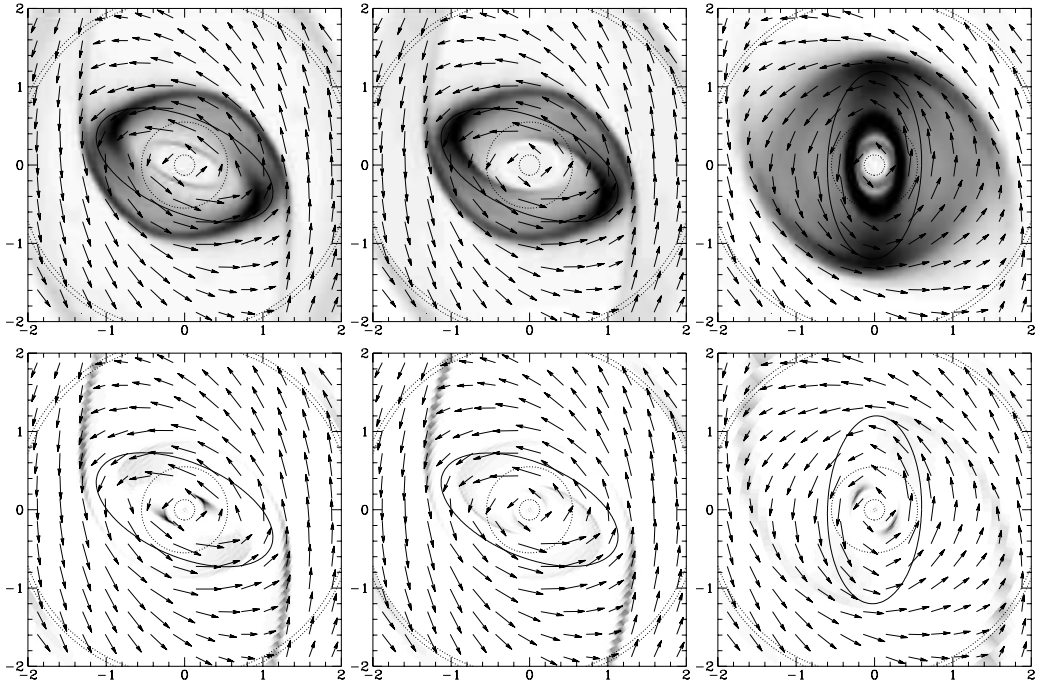}{7cm}{0}{65}{65}{-190}{-430}
\vspace{5.3cm}
\caption{Three snapshots of gas flow in the model of a doubly barred galaxy,
with density on the top and $div^2\bf v$ on the bottom. The times are
0.695 Gyr (left), 0.780 Gyr (middle), and 2 Gyr (right). The primary bar
is always vertical, and the secondary bar is outlined by the solid line.
Dotted lines indicate (innermost out) positions of the inner ILR of the
main bar,  the maximum of the $\Omega-\kappa/2$ curve at 0.55 kpc (where the
secondary bar is close to having an ILR), the outer ILR of the main bar,
and corotation of the secondary bar at 2.19 kpc.}
\end{figure}

In a doubly barred system there are loops that follow the outer bar
in its motion, and other ones that follow the inner bar. In the model
with resonant coupling (corotation of the small bar matches the outer ILR
of the main one) that we consider here, loops supporting the secondary
bar in its motion do not extend out to the outer ILR of the main bar, 
forcing the secondary bar to end well inside its corotation.
The requirement of self-consistency puts strict limits on parameters
of acceptable double bars, and rules out many hypothetical doubly-barred
systems as being far from self-consistent. It is essential to
examine gas flows in a dynamically possible model: modeling gas flows
in arbitrary potentials of double bars may be misleading, as
we do not know if such double bars can exist.

Fig.2 shows that outer regions of gas flow remain unaffected by the
introduction of the secondary bar: the principal shock retains its shape
and position. The secondary bar affects the central
regions most strongly, but no stationary straight dust lanes
form there. Rather, the flow organizes into various
elliptical and circular rings, with mass accumulating first at the
twin peaks at the ends of the secondary bar, and then settling in an
inner elliptical ring. Gas inflow to the center remains very small:
it averages about $3\times10^{-5}$ \solm/yr.

It is easy to understand this modeled flow in terms of the underlying
orbital structure. In a double bar gas stays around stable loops, and
loops supporting the secondary bar end well inside its corotation.
However, straight shocks form
only in a bar extending almost to its corotation (i.e. fast bar,
Athanassoula 1992). Therefore, what would be a straight shock in a
fast bar, here curls and forms a ring instead.  Also, no loops
corresponding to $x_2$ orbits in the secondary bar have been found,
thus one should not expect offset shocks.  Finally, loops supporting
the secondary bar originate from the $x_2$ orbits in the main bar ---
they are rather round, with no cusps, so there is no reason for shocks
in the gas flow to develop.  Thus, some self-consistent doubly barred
galaxies may lack principal shocks and dust lanes, and double bars may
not provide an efficient mechanism for fueling gas into the galactic
center.  Observations support this finding: infrared color maps of
Seyfert galaxy centers (Regan \& Mulchaey 1999) show no symmetric
straight dust lanes unrelated to the main bar, and the only doubly
barred galaxy in the sample (NGC 3081) differs from the rest by having
a central ring flattened along the secondary bar.

\section{Feeding the nucleus with a dynamically warm dissipative medium}

Regan \& Mulchaey (1999) and Martini \& Pogge (1999) report an
unexpectedly high frequency of nuclear spiral patterns in their
samples of Seyfert 2 galaxies.  Englmaier \& Shlosman (2000) found
nuclear spirals in their models for some combinations of the potential
and gas sound speed.  Their spirals are weak and can be described by a
linear wave theory.

\begin{figure}[t]
\vspace{-5cm}
\plotfiddle{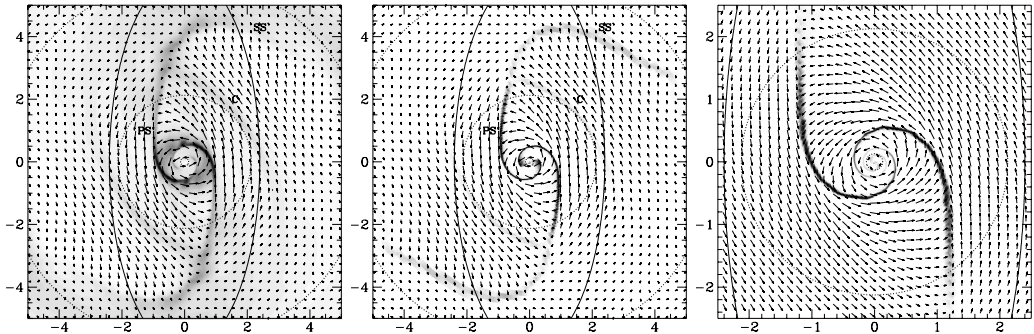}{6cm}{0}{65}{65}{-190}{-470}
\vspace{3cm}
\caption{Density (left) and $div^2{\bf v}$ (middle and right) plots of
gas flow in the inner regions of a single-bar model with gas sound speed
20 \kms. The left and middle snapshots are taken at 0.195 Gyr,
and the right one at 0.135 Gyr, just before the shock
crosses the inner grid boundary. }
\end{figure}

We modeled gas flow in the same singly-barred potential as in Fig.1,
with the sound speed in gas increased from 5 to 20 \kms.  Although the
principal shock persists, the nuclear ring is replaced by a nuclear
spiral (Fig.3), with $div^2 \bf v$ half that of the
principal shock: in this model, gas falls towards the center
in a spiral shock.  Our potential has an inner ILR, but the shock
extends further inwards, reaching the inner grid boundary at the 20 pc
radius.  The inflow to the center is very large, up to 0.15 \solm/yr,
which is enough to power a weak AGN, although the evolution after the shock
crosses the grid boundary should be interpreted with caution.

\section{Gas inflow in a weak bar}
We investigated how much the strong inflow for high sound speed
decreases if we weaken the bar. A new potential was set up, differing 
from the one above by the quadrupole moment of the bar being 10 times 
smaller, and the bar's axial ratio lowered from 2.5 to 1.5.
Although no feature similar to the straight principal shocks 
in a strong bar is present in the flow, a nuclear
spiral still forms inside the ILR (Fig.4, left).  One should expect a
flow like this, since the $x_1$ orbits supporting the weak bar in this
model are round, with no cusps, and thus do not induce shocks in gas.
Inside the ILR, as in the strong bar, gas on $x_1$ and $x_2$ orbits
interacts, giving rise to the spiral structure.  The arm-interarm
density contrast of the spiral is much lower than in the case of a
strong bar, and linear theory is sufficient to describe the gas flow:
the spiral pattern does not extend to the galaxy center (Fig.4,
center).  Adding a central black hole (or mass concentration) removes the
inner ILR, and allows the spiral pattern extend to the center (Fig.4,
right).  Surprisingly, this does not increase the inflow, which is
small in both cases --- below 0.005 \solm/yr.

\begin{figure}[t]
\vspace{-5cm}
\plotfiddle{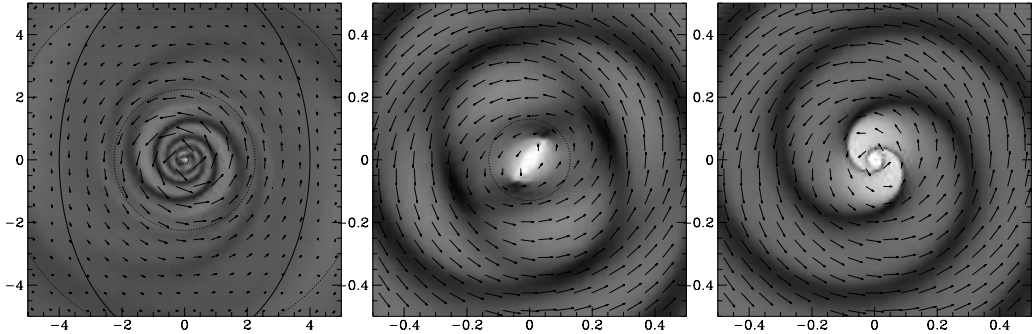}{6cm}{0}{65}{65}{-190}{-470}
\vspace{3cm}
\caption{Gas flow in the inner regions of a weak single-bar model with
density in greyscale. The left and middle plots are for the potential with
an inner ILR, the plot on the right is for a potential with a $10^8$
\solm central black hole, which removes the inner ILR. The weak bar is
outlined in solid, and dotted curves mark the same resonances as in Fig.1.}
\end{figure}

\section{Conclusions}
We have investigated basic modes of gas flow in the central parts of
galaxies by focusing on isothermal, non-selfgravitating fluid in an
external potential.  Later on, one can add other physical mechanisms
(self-gravity, star formation, magnetic fields) which will build on
this picture, and avoid confusion often caused by heuristic
explanations that include many physical mechanisms in a simplified
way.

In the fixed potential of a single bar, various modes of gas flow can
develop, some of which are able to sustain the fueling of an AGN. In
a mode with low sound speed, a single strong bar with an ILR can be
very inefficient in feeding the nucleus.  Adding an embedded secondary
bar may not increase the inflow, but if the sound speed in the nuclear
gas is high, a spiral shock develops in a single strong bar, resulting
in an inflow large enough to feed an AGN. The shock in a strong bar
corresponds to a nuclear spiral in a weak bar, but gas inflow there is
too small to feed an AGN.


\begin{references}
\reference Athanassoula E.,  1992, MNRAS, 259, 328
\reference Englmaier P., Shlosman I., 2000, ApJ, 528, 677
\reference Maciejewski W., Sparke L.S., 2000, MNRAS, 313, 745
\reference Maciejewski W., Teuben P., Sparke L., Stone, J., 2001, 
MNRAS, submitted
\reference Martini P., Pogge R.W., 1999, AJ, 118, 2646
\reference Piner B.G., Stone J.M., Teuben P.J., 1995, ApJ, 449, 508
\reference Regan M.W., Mulchaey J.S., 1999, AJ, 117, 2676
\reference Shlosman I., Frank J., Begelman M., 1989, Nature, 338, 45
\end{references}
\end{document}